# COMBINING ADVANCED VISUALIZATION AND AUTOMATIZED REASONING FOR WEBOMETRICS: A TEST STUDY

CLAIRE FRANCOIS[1]
JEAN-CHARLES LAMIREL[2]
SHADI AL SHEHABI[2]

1 INIST - CNRS, 2 allée du Parc de Brabois
F-54514 Vandoeuvre-lès-Nnacy Cedex - France
claire.francois@inist.fr
2 LORIA
B.P. 239, 54506 Vandoeuvre-lès-Nancy Cedex - France
lamirel@loria.fr, alshahab@loria.fr

## *Abstract*

This paper presents a first attempt for performing a precise and automatic identification of the linking behaviour in a scientific domain through the analysis of the communication of the related academic institutions on the web. The proposed approach is based on the paradigm of multiple viewpoint data analysis (MVDA) than can be fruitfully exploited to highlight relationships between data, like websites, carrying several kinds of description. It uses the MultiSOM clustering and mapping method. The domain that has been chosen for this study is the domain of Computer Science in Germany. The analysis is conduced on a set of 438 websites of this domain using all together, thematic, geographic and linking information. It highlights interesting results concerning both global and local linking behaviour.

## 1. *Introduction*

The use of links between academic websites to create an informal mode of scholarly communication is a promising new field of investigation. Hence, thanks to qualitative analyses of links (Wilkinson *et al.* 2003), the metrics based upon link counts can be seen to be measuring an agglomeration of connections related to scholarly activities in a wide variety of ways. However, thanks to others analyses (Chu 2005), the evaluative link studies should not only consider link counts but also reasons for linking to ensure the validity of such research. Moreover, the linking behaviour has been highlighted as dependant of the studied domain. As an example, (Thelwall *et al.* 2003) found that the mathematics and computer sciences domains are more interlinked than other scientific domains. Several recent works have gone one step further, by demonstrating that a precise analysis of linking behaviour depends on the capability of taking into account several complementary factors, like discipline and geographic factors (Thelwall 2002). Nevertheless, performing such a kind of analysis at a large scale prohibits the use of a manual mode, as it has been proposed up to now.

Graph theoretic methods (Bjöneborn & Ingwersen 2004, Thelwall *et al.* 2005) or standard neural methods (Decker 2001) have been commonly used for automatically investigating link structures on the Web. Their main defect is that they do not represent a reliable support for performing an accurate analysis based on multiple factors.

The paradigm of multiple viewpoint data analysis (MVDA) (Lamirel 1995) represents a promising basis to solve the problem of automatic analysis of linking behaviour. Hence, it can be fruitfully exploited to highlight relationships between data coming from different types of source. It can also be used to perform cross-analysis between data belonging to the same dataset, using different description criteria. In the domain of Webometrics, MVDA has been already successfully used to establish the research policy and the industrial relationships of a well-known institution by automatically crossing content and outlinks analyses performed on the same reference set of webpages extracted from the institution website (Lamirel *et al.* 2004). The role of the present study is to go one step further. It represents a first attempt of identification of the linking behaviour in a scientific domain through the analysis of the communication of the related academic institutions on the web using a MVDA approach based on clustering and mapping method. The domain that has been chosen for this study is the domain of Computer Science in Germany.

Section 2 of the paper proposes an overall presentation of the approach. Section 3 describes the preparation of the dataset merging data about European universities and research institutes websites collected in the framework of the IST-EISCTES project. Section 4 presents the results of link behaviour analyses performed on the test dataset by the use of the MultiSOM approach: both global and local linking behaviour analyses are considered. Section 5. Conclusions are finally presented.

## *2. The MultiSOM model*

The basic principle of the self-organizing maps (SOM) is that our knowledge organization at higher levels is created during learning by algorithms that promote self-organization in a spatial order (Hinton 1989, Kaski *et al.* 1998, Kohonen 1990). Thus, the architecture form of the SOM network is based on the understanding that the representation of data features might assume the form of a self-organizing feature map that is geometrically organized as a grid or lattice. The SOM algorithm takes thus a set of N-dimensional data as input and maps them onto nodes of a two-dimensional grid, resulting in an orderly feature map (Kohonen 1990). The communication between self-organizing maps has been firstly introduced in the context information retrieval for analyzing the relevance user’s queries regarding the documentary database contents (Lamirel 1995). The resulting model, so called MultiSOM, represents both a major amelioration of the basic Kohonen SOM model and the first operational implementation of the multiple viewpoint data analysis (MVDA) paradigm.

The viewpoint building principle consists in separating the description space of the data into different subspaces corresponding to different criteria of analysis. The set of $V$ all possible viewpoints issued from the description space $D$ of a dataset can be defined as:

$$V = \{v_1, v_2, \ldots, v_n\},\ v_i \in P(D)$$

where each $v_i$ represents a viewpoint and $P(D)$ represents the set of the parts of the description space of the data $D$.

Three main remarks follow the above definition: (1) the viewpoint subsets issued from $V$ may overlap one to another; (2) the union of the different viewpoints can be viewed as the overall description space of the data; (3) the most suitable basis an for homogeneous management of the viewpoints is a vectorial description space. As an example, an image can be simultaneously described using 3 different viewpoints represented by: (1) a key-term vector; (2) color histogram vector; (3) a feature vector.

The principle of the MultiSOM model is to be constituted by several SOM maps that have been generated from the same data. Each map is itself issued from a specific viewpoint. The relation between maps is established through the use of one main communication mechanism. The inter-map communication mechanism enables to highlight semantic relationships between different topics (i.e. clusters) belonging to different viewpoints related to the same data. In MultiSOM, this communication is based on the use of the data that have been projected onto each map as intermediary nodes or activity transmitters between maps (see Figure 1).

Figure 1: Inter-map communication principle

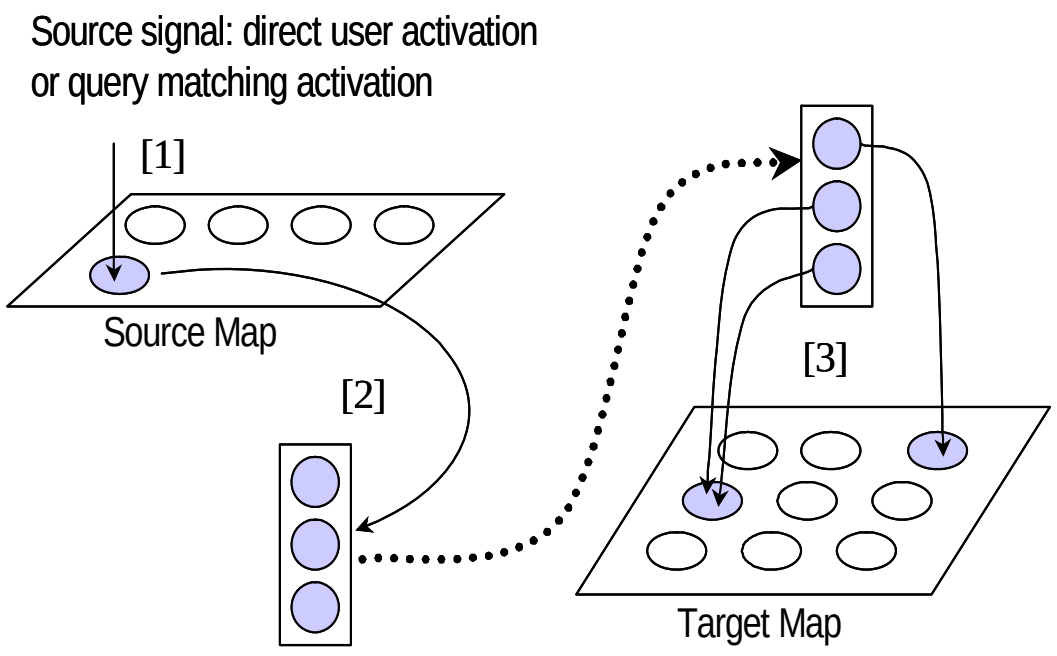

The inter-map communication is established by standard Bayesian inference network propagation algorithm which is used to compute the posterior probabilities of target map's node $T_k$ which inherited of the activity (evidence $Q$) transmitted by its associated data nodes. This computation can be carried out efficiently because of the specific Bayesian inference network topology that can be associated to the MultiSOM model. Hence, it is possible to compute the probability $P(act_m|T_k,Q)$ for an activity of modality $act_m$ on a target map node $T_k$ which is inherited from activities generated on the source map. This computation is achieved as follows (Al Shehabi & Lamirel. 2004):

$$P(act_m|T_k,Q) = \frac{\sum_{d \in act_m, T_k} Sim(d,S_d)}{\sum_{d \in T_k} Sim(d,S_d)}$$

such that $S_d$ is the source node to which the data $d$ has been associated, $Sim(d, S_d)$ is the cosine correlation measure between the description vector of the data $d$ and the one of its source node $S_d$ and $d \in act_m, T_k$ if it has been activated with the modality $act_m$ from the source map.

After each map building, the main characteristics of the nodes, that represent the clusters resulting from the topographical clustering process, have to be highlighted to the analyst in order to provide him an overview (i.e. a global summary) of the analysis results. This task is difficult because the description vectors of the obtained clusters consist in complex weighted combination of indexes extracted from the data. A first set of methods for solving this problem is presented in Lamirel *et al.* (2000) and Polanco *et al.* (2001). It consists of cluster labelling and map zoning methods, dividing a map into homogeneous information areas. Figures 3 and 4 of section 4 illustrate the overall results of these methods.

## 3. *Data preparation*

The data used in our experiment are information about laboratory websites of the European Universities have been collected in the EICSTES project (http://www.eicstes.org/). These data are more extensively described by Arroyo *et al.* 2003. They have originally the form of multiple tables issued from a preliminary basic structural analysis of the websites. Thus, a first phase of our data preparation consists in obtaining an overall description of each website by merging all the individual description tables into a global description table. After this first merging, the global description of a website includes theURL, the name of the organization and department, the number of pages, the number of outgoing links, the number of repeated outgoing links, the Geographic situation : country and town code, country and town name, the Domain: code, label and related domain codes, the Inlinks: list of incoming links with their URLs and the number of links coming from these URLs, the Outlinks: list of outgoing links with their URLs and the number of links going to these URLs.

The investigated data set covers the websites of the research laboratories in the 15 countries of the European Union before March 2004. The domain categorization of the websites is based on the UNESCO classification. The UNESCO code is a classification that is used to allocate a scientific domain to a website starting from its content. As the original data is too general, we have firstly decided to focus our study on a specific thematic domain. The 1203 UNESCO code that deals in a global way with computer science is used for websites selection. In a second step, we have chosen to focus on German research laboratories' websites as reference websites for our study. The kernel set of 438 websites is selected in this way. Hence, the study will more precisely focus on the relationships existing between German universities relatively to a European context. The Outlinks initially represented the overall set of links emitted by the websites of the kernel set. For the purpose of our study, we have restricted this set to the subset of links emitted in the direction of a website belonging to the investigated data set, that is, the websites of the research institutes in the 15 countries of the European Union. In a complementary way, the Inlinks are calculated starting from the set of websites of the research institutes in the 15 countries of the European Union.

**Viewpoints definition**

Starting from the websites characteristics, viewpoints are defined that describe their distribution in various thematic sub-domains (1), their geographical repartition (2), and their network of Outlinks (3) and Inlinks (4).

Each viewpoint is represented in the form of a matrix including websites codes in its rows and the description criteria associated to the viewpoint in its columns. Thus, the table 1 highlights that the kernel set is indexed by 96 towns (German cities), 93 thematic sub-domains (Unesco codes), 2079 Outlinks and 2839 Inlinks, respectively. Looking to these values, a first general remark can be made: German research laboratories are significantly more cited (2839 Inlinks) than they cite (2079 Outlinks). The matrices listed in table 1 represent the direct entry data of the clustering MultiSOM application.

Table 1: Matrix size for each viewpoint

| Viewpoints | Matrix size |
|---|---|
| Towns (1) | 438 * 96 |
| Thematic sub-domains (2) | 438 * 93 |
| Outlinks (3) | 386 * 2079 |
| Inlinks (4) | 388* 2839 |

## 4. *Data analysis*

### 4.1. *Clustering and mapping using SOM*

A map is computed for each viewpoint. In order to define the optimum size of that map, different square maps starting from 9 nodes (3*3) to 400 nodes (20*20) are calculated using the SOM basic clustering application "SOM_PACK" (*SOM papers)*. The choice of the best map is based on an optimisation algorithm using specific quality criteria (recall, precision and F-measure) derived both from information retrieval and symbolic learning. This approach is more extensively described in Lamirel *et al.* (2004b). Table 2 presents the final results of the whole map construction process, the optimum number of clusters and the quality values (recall, precision and F-measure) for each viewpoint. As defined in Lamirel *et al*. (2004), the quality values range between 0 and 1, 1 representing the best possible value in all cases.

Table 2: Optimum number of clusters for each viewpoint

| Viewpoints | Number of clusters | Recall | Precision | F_measure |
|---|---|---|---|---|
| Towns (1) | 256 | 1,00 | 0,90 | 0,95 |
| Thematic sub-domains (2) | 81 | 0,74 | 0,72 | 0,73 |
| Outlinks (3) | 144 | 0,49 | 0,49 | 0,49 |
| Inlinks (4) | 289 | 0,58 | 0,51 | 0,54 |

Table 2 highlights very high quality values for Towns and Sub-domains viewpoints, and conversely, quite low quality values for the Outlinks and Inlinks viewpoints. Hence, in the case of the Towns and Sub-domains viewpoints, clusters are quite homogeneous and distinct one to another. This distribution is carried out easily insofar as each website is indexed by a low number of weakly overlapping properties. As soon as each website presents a relatively significant number of incoming and outgoing links, overlaps are thus potentially much more significant, this implies relatively moderate quality values for the Outlinks and Inlinks viewpoints, even after the optimisation process. These preliminary results will be taken into account in the remaining part of our study.

## *4.2. Description of the maps*

For the viewpoint (1), the map clusters gather websites sharing their geographic location. For the viewpoint (2), the map clusters gather websites sharing their overall research profile (i.e. combination of Unesco codes). For the viewpoint (3), the map clusters gather websites sharing their Outlinks: they are described by the targets of the links. The viewpoint (4) is the equivalent of (3) using the Inlinks: the maps clusters are described by the targets of the links.

The easiness of interpretation of a map not only depends on the map quality (see section 4.1) but also on complementary factors, like the granularity of description. Two typical cases of maps are described hereafter.

Figure 2: The Sub-domains map

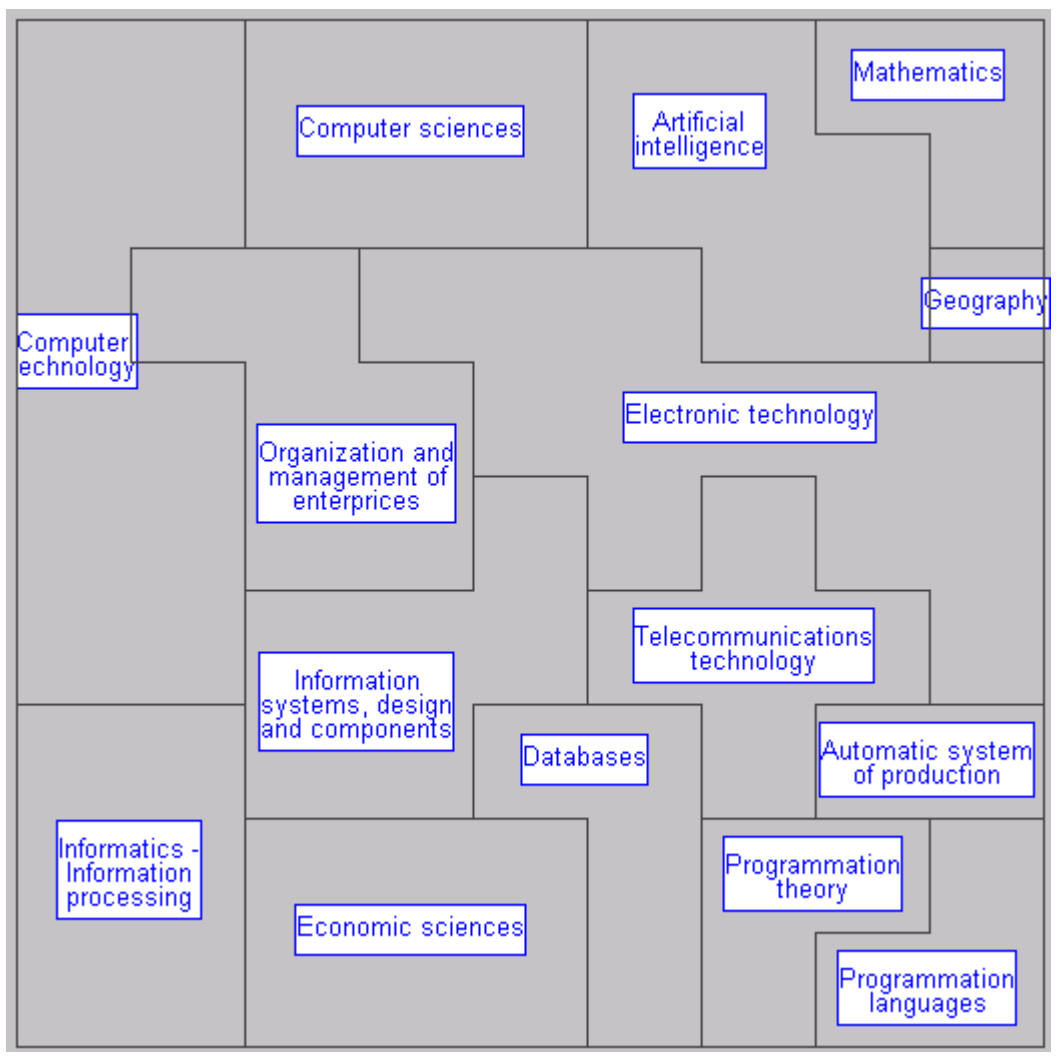

Figure 3: The Outlinks map

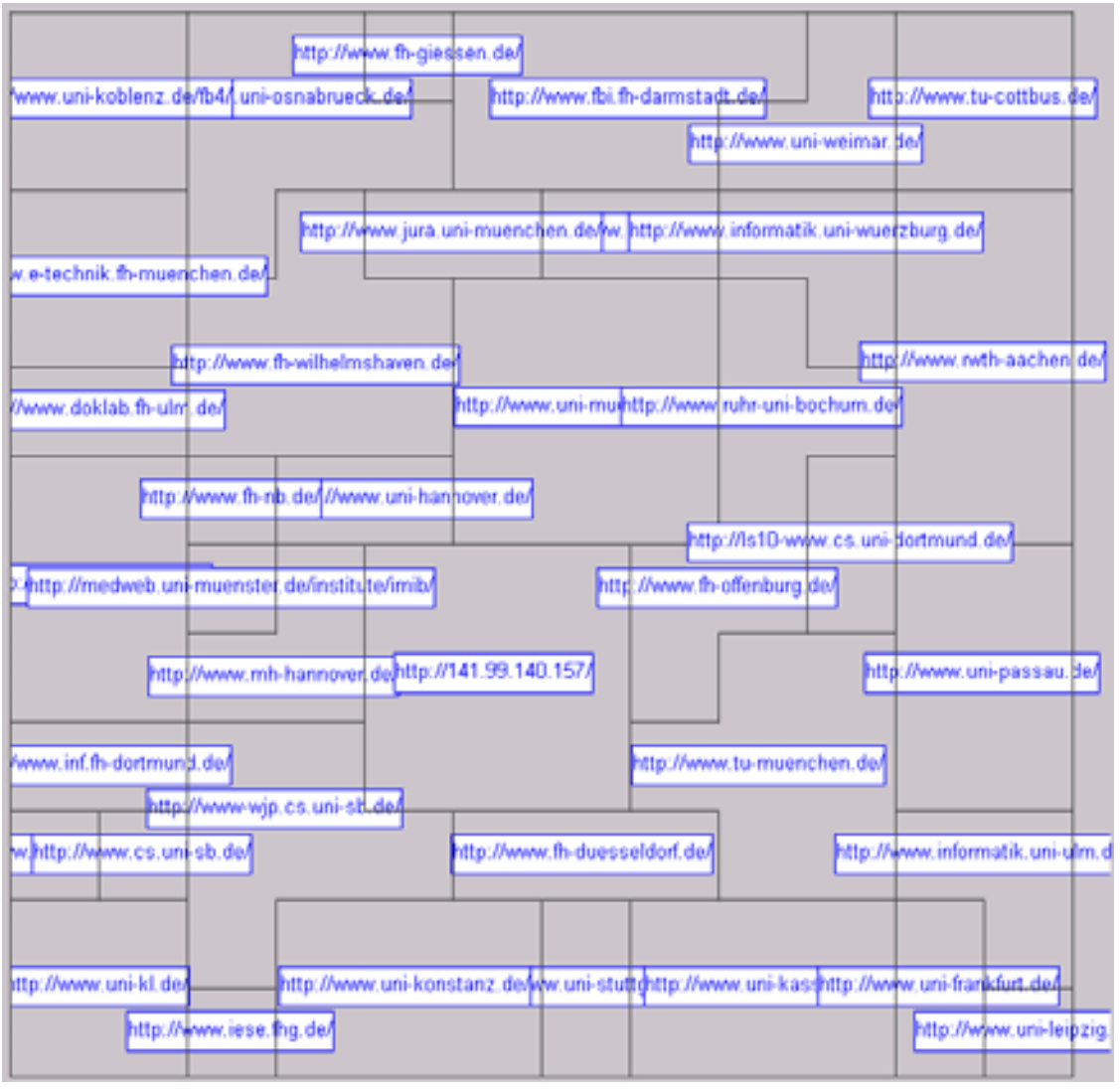

Case 1: The Sub-domains map (figure 2): it is a relatively small map (81 clusters) with relatively large information areas. The map clusters have generic headings and some of them contain a large number of websites. For example, five clusters (among a total of 81) gather 227 websites (among a total of 438). A more accurate indexation would have allowed a better distribution of the websites in clusters of more homogeneous size, and consequently, a more precise map content interpretation.

Case 2: The Outlinks map (figure 3): each website has many Outlinks leading to high granularity of websites Outlinks descriptions, and thus, to potentially more interesting map configuration. Thus, the Outlinks map has information areas of relatively homogeneous and small size. Such a map configuration highlights a homogeneous website outlinking behaviourConsistency measure between the different viewpoints

## *4.3. Consistency measure between the different viewpoints*

The propagation consistency measure is presented in Al Shehabi & Lamirel (2005). It evaluates the activity focalization generated by a source map *S* on a target map *T*. It is a non necessary symmetrical measure. The propagation coherency (PC) is given by:

$$PC = \frac{1}{|\overline{S}|} \sum_{k;S_k \neq \phi} \frac{\sum_i P(act \,|T_{i_k})}{D_k + 1}$$

where

$$D_k = \frac{2\sum_i \sum_{j=i+1} \|T_{i_k} - T_{j_k}\|}{|T^k|.(|T^k| - 1)}$$

and $\overline{S}$ represents the peculiar set of nodes extracted from the source map of *S*, which verifies:

$$\overline{S} = \{S_k \in S \,|S_k \neq \phi, S_k \in act\}$$

$T^k$ is the set of the activated nodes in the target map through the activation of the source node $S_k$

$$T_k = \{T_{ik} \in T | T_{ik} \in PRG(S_k)\}$$

Each node in the SOM grid is represented its coordinate position

$$T_{i_k} = (a^{i_k}, b^{i_k})$$

In a practical way, the propagation consistency takes into account the focalization of the activity generated by the clusters of the source map on a target map (figure 4). A strong focalization of all the clusters of a source map on a target map will lead to a high consistency.

Figure 4: Focalization of the activity propagation

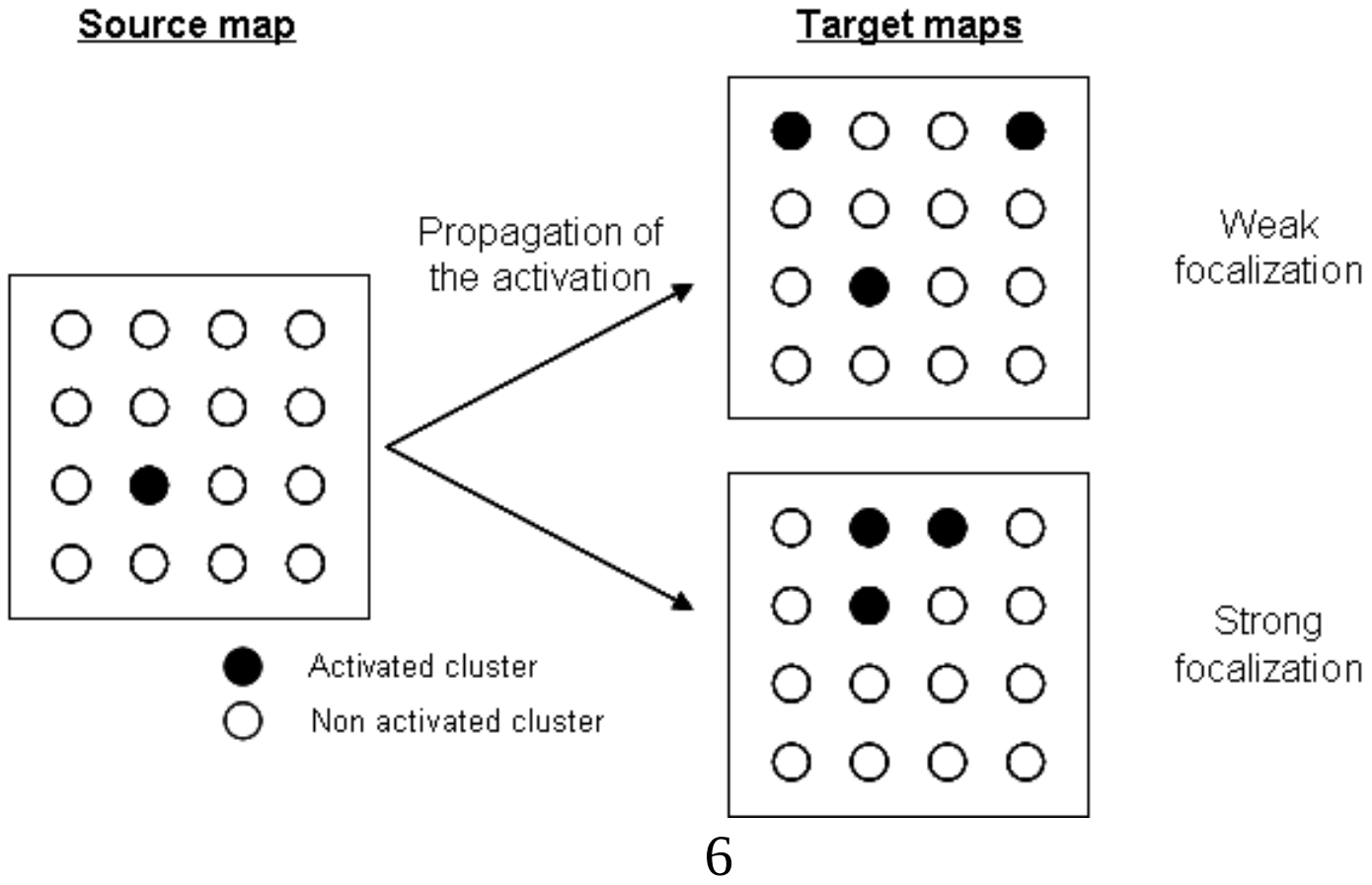

A high consistency measured from a source map to a target map means that the clusters of the target map can be well explained by the clusters of the source map. Please remember that consistency is a non symmetrical measure. Thus, if the reverse consistency, measured from the target map to the source map, is also high, this can lead to consider that the two maps are similar. On the contrary, if the reverse consistency is low, there is no evidence concerning any converse explanation provided by the target map on the source map.

Table 3 presents the consistencies values existing between the various viewpoints. The results being asymmetrical, the table must be read lines towards column as shown by the arrow. The values lie between 0 and 1 out of the diagonal, while the unity value can be observed in the diagonal, a map being perfectly self-consistent.

Table 3: Consistency values between the maps of the different viewpoints

| | **Towns** | **Sub-domains** | **Outlinks** | **Inlinks** |
|---|---|---|---|---|
| **Towns** | 1 | 0.096 | 0.250 | 0.270 |
| **Sub-domains** | 0.216 | 1 | 0.554 | 0.416 |
| **Outlinks** | 0.117 | 0.095 | 1 | 0.270 |
| **Inlinks** | 0.080 | 0.055 | 0.130 | 1 |

**Behaviour of linking according to the research sub-domains**

The highest value of consistency is Sub-domains towards Outlinks (0.554). This can be interpreted as an indicator of coherence of outlinking behaviour of the laboratories belonging to common research sub-domains. The consistency value of Sub-domains towards Inlinks is significantly high (0.416), while weaker than the former value. This indicates a good identification of the research sub-domains of the laboratories, although they don't have the control of the links they receive.

In a reverse way, the consistency value of Outlinks towards Sub-domains is particularly low (0.095), as well as the consistency value of Inlinks towards Sub-domains (0.055). These values indicate that the global linking behaviour of the research laboratories largely overcome their research sub-domains.

**Geographical distribution of research sub-domains**

The value of consistency of Sub-domains towards Towns (0.216) is moderately high, indicating certain coherence in geographical distribution of the research sub-domains: research sub-domains tend to gather in specific cities. Conversely, the value of consistency of Towns towards Sub-domains is low (0.096), indicating that various research sub-domains coexist in the German cities.

**Overall comparison of inlinking and outlinking behaviours**

Table 3 shows that the values of consistency of Outlinks towards all viewpoints are higher than those of Inlinks. This indicates that the clusters of the Outlinks map are more coherent than those of the Inlinks map. We can generalize this observation by proposing the assumption that the outlinking behaviour of a research laboratory is more coherent than its received referencing. Nevertheless, this observation should be taken with care, as soon as Inlinks and Outlinks maps quality are moderate.

The former discussion highlights that the analysis of the consistency in multi-viewpoints context of the MultiSOM approach represent a sound basis to get an overall view of the linking behaviour in a large scale context. More focused strategies based on the same approach are presented hereafter. As it will be shown, they can be used for illustrating local phenomenon related to linking behaviour.

## *4.4. Local analysis using the inter-map communication mechanism*

The MultiSOM inter-map communication mechanism can be used in an interactive mode to highlight specific relationships between clusters of different maps. For this purpose, an activity is assigned to a cluster, or to an information area, of a source map. Then, the mechanism of propagation of the activity to all the other maps (target maps) is applied (figure 1). Finally, the focalization profile of the target maps is analyzed (figure 4). Said profile indicates a more or less great coherence between activated cluster, or area, of the source map and the activated clusters of the target maps.

We present hereafter two examples of application of such a mechanism for performing local analysis of link behaviour. The first example is an analysis of the link behaviour of the research laboratories localised in a town. For that analysis, we have chosen the biggest town cluster considering the number of associated websites, i.e. the Munich town cluster. The second example is a study of the link behaviour of the laboratories sharing the same research sub-domain. In order to observe a potential focalization, we have chosen a sub-domain which is not too general, like Computer Sciences or Computer Technology. Thus, we have focused the study on the Mathematics sub-domain.

**Example of Munich**

The figure 5 illustrates the overall strategy that has been used for realizing the Munich study. In a first step, a source activity is set on the Munich information area of the Town map and the resulting activation spreading on the Outlinks map is analysed. In a second step, the overall zone of the Outlinks map on which the activity has focalized at the first step is reused as an activity source, and the backward activity spreading on the Town map is analyzed.

Figure 5: Inter-map communication mechanism starting from the Munich cluster
Step 1: communication from Towns map towards Outlinks map,
Step 2: communication from Outlinks map towards Towns map.

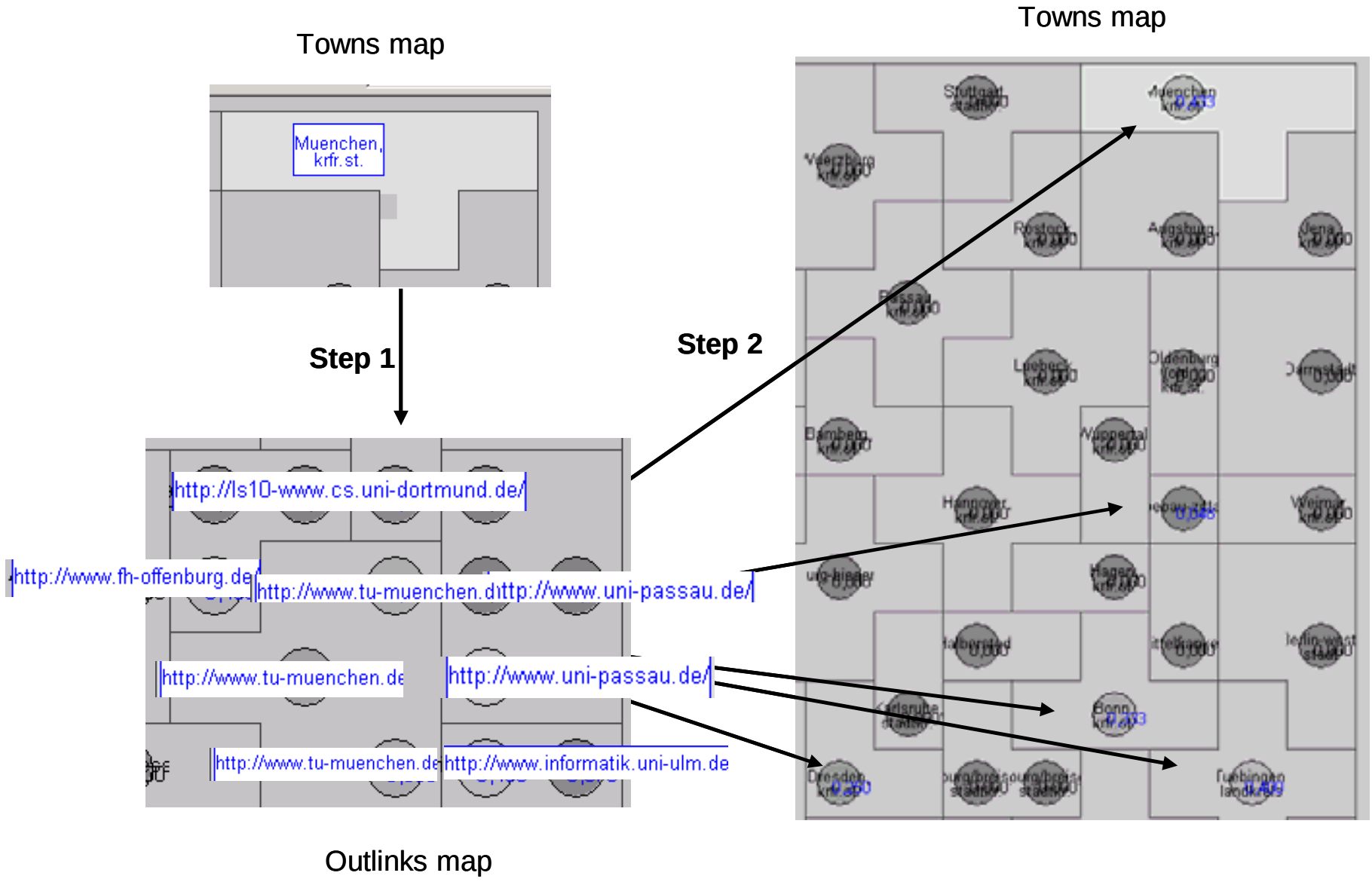

**Step 1:** the propagation of the activity starting from the Munich information area and going towards the Outlink map concentrates around an information area, which gathers 3 clusters whose profile is dominated by the URL http://www.tu-muenchen.de/ (figure 5). The activated clusters located around this information area have the following dominant URLs in their profile:

http://www.uni-passau.de/

http://www.informatik.uni-ulm.de/

http://www.fh-offenburg.de/

http://ls10-www.cs.uni-dortmund.de/

The above mentioned URLs correspond to main websites cited by the websites of Munich laboratories. They thus summarize the outlinking behaviour of these latter laboratories. This led us to conclude to a relatively local outlinking behaviour of the Munich laboratories, i.e. referecing towns mostly located in the South of Germany (Passau, Ulm, Offenburg).

**Step 2:** the "http://www.tu-muenchen.de/" information area of the Outlinks map gathers websites having this URL as the dominant URL in their profile. An activity propagation starting from this information area is launched towards the Towns map. Its role is to highlight the geographical spreading of the laboratories sharing their outlinking profile with the dominant outlinking profile of the Munich laboratories. This activity propagation results in activated clusters all over the Towns map (Munich, Bonn, Tuebigen, Dresden, and Loebau-Zittau). This led us to conclude that: even if the linking behaviour of Munich is relatively local (see step 1), Munich's type of outlinking is not specific to Munich and/or to its direct surrounding. Thus, it enables us to reject the "closed world" outlinking hypothesis concerning Munich.

**Example of Mathematics**

The figure 6 illustrates the overall strategy that has been used for realizing the Mathematics study. This strategy is similar to the one that has been used for the Munich study. In a first step, a source activity is set on the Mathematics information area of the Sub-domains map and the resulting activation spreading on the Outlinks map is analysed. In a second step, the overall zone of the Outlinks map on which the activity has focalized at the first step is reused as an activity source, and the backward activity spreading on the Sub-domains map is analyzed.

Figure 6: Inter-map communication mechanism starting from the Mathematicsinformation area

Step 1: communication from Sub-domains map towards Outlinks map,
Step 2: communication from Outlinks map towards Sub-domains map.

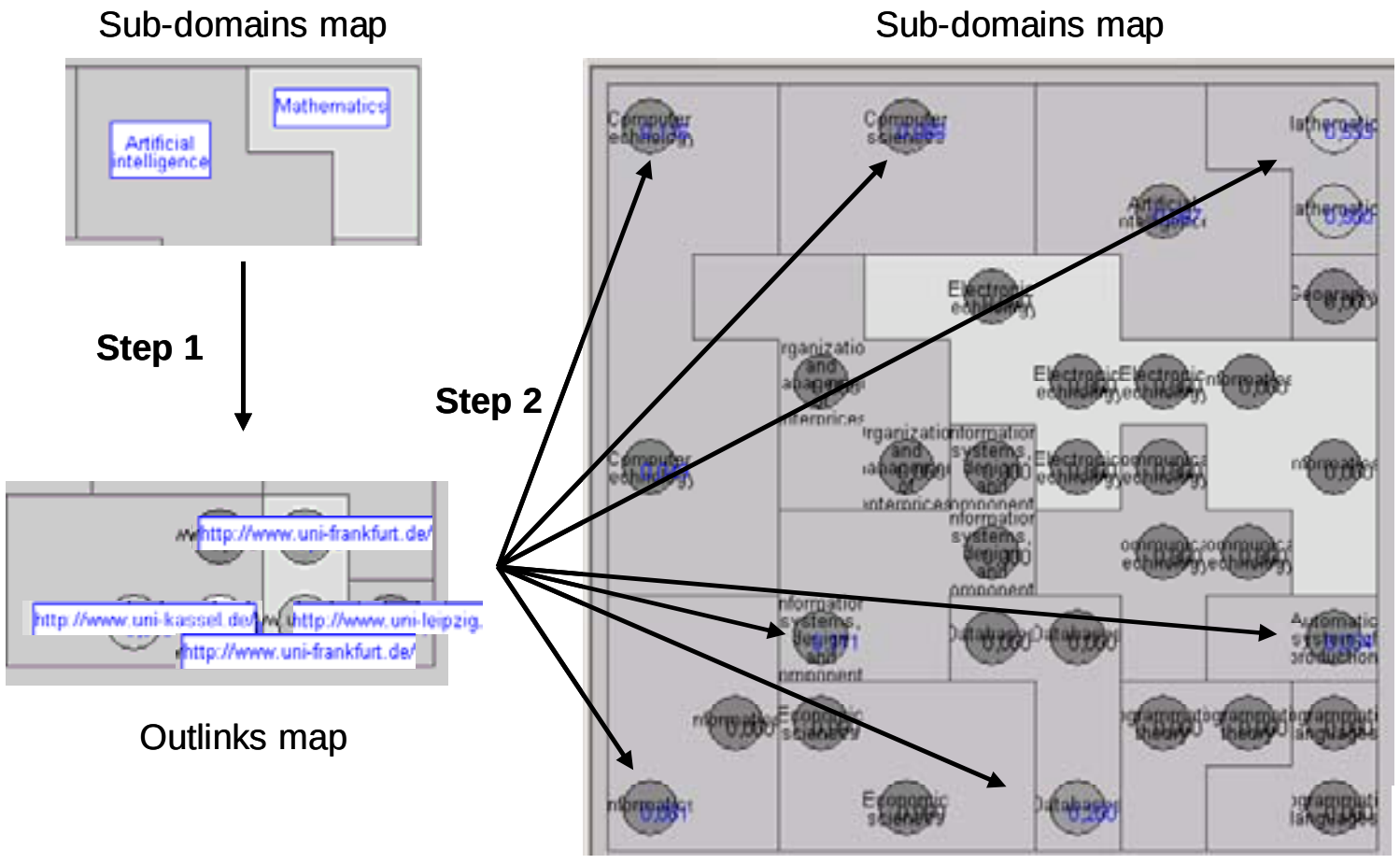

**Step 1:** the propagation of the activity starting from the Mathematics information area towards the Outlink map highly focuses around an information area, which gathers 3 clusters whose profile is dominated by the URLs:

http://www.uni-kassel.de/,

http://www.uni-frankfurt.de/,

http://www.uni-leipzig.de/.

These URLs correspond to main websites cited by the websites of the Mathematics information area. They thus summarize the outlinking behaviour of these latter websites. As the activity is focalized on the target map, it also means that this outlinking behaviour is specifically coherent. This local observation is compliant with the high value of the overall consistency of Subdomains relatively to Outlinks (0,554) presented in table 3.

**Step 2:** an activity propagation starting from the clusters whose dominant URLs have been highlighted at step 1 (http://www.uni-kassel.de/, http://www.uni-frankfurt.de/, http://www.uni-leipzig.de/) is launched towards the Sub-domains map. Its role is to highlight the thematic spreading of the laboratories sharing their outlinking profile with the dominant outlinking profiles of the Mathematics domain. This activity propagation results in activated clusters all over the Sub-domains map (Mathematics, Artificial intelligence, Databases, Programmation theory, Automatic system of production, Informatics - Information processing, Information systems, design and components, Computer technology, Computer sciences). This local observation is compliant with very low value of the overall consistency of Outlinks relatively to Subdomains (0,095) presented in table 3. This led us to conclude that: even if the linking behaviour of the Mathematics is highly coherent (see step 1), Mathematics's type of outlinking is not specific to this research domain.

The former discussion highlights that the local focalization analysis can be successfully used for illustrating local phenomenon related to linking behaviour. In a complementary way, they could also be used for highlighting local phenomenon whose behaviour significantly differs from the global behaviour measured by the consistency analysis. This mechanism would then be a precious one for highlighting exceptions.

## *5. Conclusion*

This paper presents a first attempt for performing a precise and automatic identification of the linking behaviour in a scientific domain through the analysis of the communication of the related academic institutions on the web. Our approach is based on the paradigm of multiple viewpoint data analysis (MVDA). This paradigm has been exploited to carry out a complete analysis of a website dataset through the use of multiple kinds of website descriptions: geographical localisation, research topics, and networks of links between websites. Our dataset covers 438 websites of German Computer Science research laboratories.

Both local and global link behaviour analyses have been carried out using the MultiSOM approach. MultiSOM is a clustering and mapping approach aiming at gathering data into homogeneous subsets projected on two dimensional maps. It is associated with an original mechanism of communication between maps associated to different viewpoints. This mechanism has been used in an automatic mode in order to highlight general information related to the link behaviour. A local link behaviour analysis has been achieved by focusing on data subsets associated map clusters and by using the interactive form of that mechanism.

Several preliminary results concerning the linking behaviour of the German research laboratories have been obtained. One first result is the thematic consistence of both the outlinking and the inlinking behaviour. A similar phenomenon has been observed to a less extent concerning geographic consistence of the research sub-domains. A general assumption has also been made by the observation of maps consistency. It is related to the fact that the outlinking behaviour of a research laboratory is more coherent than its received referencing. We must note that other experiments related to datasets issued from different countries must be done to validate these results.

Moreover, a regional outlinking policy has been highlighted focusing the analysis on a specific German city although this experiment must be completed to generalize this observation to all the cities of the dataset.

A local analysis of websites subsets associated either to a city or to a sub-domain provides the analyst with a clear overview on neighbourhood of the elements of a sub-network.

The main strength of our approach is to dynamically combine multiple kinds of data descriptions for performing automatized link analyses. Nevertheless, some potential correspondences regarding the behaviour of this approach must be more in deeply investigated: as an example, the relation between the consistencies values between the map and the local analysis using these different maps. This knowledge will certainly help us to validate important hypotheses. For that purpose, we plan to achieve methodological experiments on different dataset in the near future.